\documentclass[%
 reprint,
superscriptaddress,
 amsmath,amssymb,
 aps,
prb,
]{revtex4-2}
\usepackage{amsmath, bm, physics,amssymb}
\usepackage{color}
\usepackage{graphicx}
\usepackage{siunitx}
\graphicspath{{picFolder/}}

\usepackage[whole]{bxcjkjatype} 
\usepackage[colorlinks=true,urlcolor=blue,citecolor=blue,linkcolor=blue,breaklinks=true]{hyperref}
\input glyphtounicode
\begin{document}
\title{Spinon shift current in a noncentrosymmetric quantum spin chain}
\author{Ryosuke Yamashita}
\affiliation{Department of Physics, Kyoto University, Kyoto,
  606-8502, Japan}
\author{Shintaro Takayoshi}
\affiliation{Department of Physics, Konan University, Kobe 658-8501, Japan}
\author{Takahiro Morimoto}
\affiliation{Department of Physics, Kyoto University, Kyoto,
  606-8502, Japan}
\date{\today}
\begin{abstract}
We theoretically study direct current generation in a quantum spin chain induced by spinon excitations by light irradiation.
We consider a $S=1/2$ one-dimensional (1D) antiferromagnetic XXZ model with magnetoelectric coupling that describes multiferroics with broken inversion symmetry.
We perform the real-time simulation using infinite time-evolving block decimation, and demonstrate the direct current generation under light irradiation. 
By comparing the second order nonlinear conductivity and the two-spinon excitation spectra of a 1D XXZ model, we confirm that the spinon excitations are the origin for the direct current generation in the quantum spin chain.
We find that the bulk photovoltaic effect is driven by electric polarization carried by the spinons through the shift current mechanism, and thus is regarded as ``the spinon shift current.''
\end{abstract}
\maketitle

\section{Introduction}
Nonlinear responses in solids have attracted significant attention in both their fundamental and their application aspects \cite{Boyd,Bloembergen,Sturman}.
For example, they include important functionalities of quantum materials such as optical rectification, high-harmonic generation (HHG), and nonreciprocal transport. 
Among these phenomena, the bulk photovoltaic effect (BPVE) arising from second-order nonlinear responses
\begin{equation}
    j^{(2)}_\mathrm{DC}\equiv\sigma^{(2)}_\mathrm{DC}E(\omega)E(-\omega)
\end{equation}
has importance on its application to photoelectric conversion \cite{ma2021topology,orenstein2021topology,dai2022recent,ma2023photocurrent,Morimoto-JPSJ23}. 
In particular, "shift current" is one mechanism for the BPVE in noncentrosymmetric crystals, which originates from the shift of the electron wave packet during the optical transition \cite{Baltz,Sipe,Young-Rappe,doi:10.1126/sciadv.1501524,Resta24}.
Namely, the optical transition produces electron-hole pairs with a finite electric polarization, resulting in DC generation in the steady state. 
This shift of the wave packet is characterized by a geometrical quantity called the shift vector and is closely related to the modern theory of electric polarization.

Since the shift current is essentially driven by the electric polarization of excitations, it is expected that the shift current mechanism is also applicable to correlated electron systems. 
Indeed, recent studies reveal that shift current responses can be mediated by quasiparticles and collective excitations in interacting electron systems, as exemplified by BPVEs due to magnons \cite{PhysRevB.100.235138}, phonons \cite{doi:10.1073/pnas.2122313119,PhysRevB.110.045129,p4s2-jpyk}, and excitons \cite{PhysRevB.94.035117}. 
Among these excitation-mediated optical responses, the photocurrent generations via spin excitations are of particular interest due to their connection to multiferroics; these are strongly correlated materials where spin dynamics induce a polarization response through magnetoelectric cross-correlations \cite{Fiebig_2005,Mostovoy,Tokura_2014}. It was shown that spin excitations in multiferroic materials can give rise to the photovoltaic effect owing to the shift current mechanism, whose energy range is much lower than the conventional visible light regime and typically falls into the terahertz frequency range \cite{Ogino2024}. 
This suggests that multiferroic materials are a promising platform for novel photoelectric conversion functionalities within the terahertz regime.

Meanwhile, spin excitations in quantum spin systems have long been a central problem in strongly correlated electron systems \cite{Fradkin_2013,Sachdev_2011}. 
In quantum spin systems, there emerge exotic excitations, as well as exotic phases, that cannot be found in classical spin systems. The most famous ones are spinon excitations in spin liquid phases \cite{Savary_2017,broholm2020quantum,RevModPhys.89.025003}.
For example, a spinon's contribution to the linear optical response \cite{PhysRevLett.103.177402} and nonlinear spin current \cite{PhysRevLett.122.197702} has been investigated in multiferroic quantum spin chains previously.
Therefore, it is interesting to ask how those quantum spin excitations affect the nonlinear optical responses and especially contribute to the shift current response under strong spin fluctuations near quantum phase transition.

Thus, our primary objective is to capture the contribution of quantum spin excitation in the BPVE. To this end, we need to employ the nonpertubative way which involves the quantum correlation effect. 
Also, it is desirable for the method to be well suited for real-time evolution of many-body systems. The infinite time-evolving block decimation (iTEBD) method \cite{PhysRevLett.98.070201}, a numerical technique of 1D quantum many-body systems based on the tensor network formalism \cite{RevModPhys.93.045003}, is known to allow for a precise treatment of quantum many-body effects not only in equilibrium but also in nonequilibrium states \cite{PhysRevB.99.184303,PhysRevB.103.035110,murakami2022exploring,PhysRevB.105.L241108,PhysRevLett.134.096504} and, therefore, is well suited to our purposes. 

In this paper, we perform a theoretical analysis of photocurrent generation induced by quantum spin excitations in a 1D magnetic insulator. 
Specifically, we construct a 1D spin model based on the XXZ model by incorporating a magnetoelectric coupling known as the exchange-striction mechanism. 
We perform iTEBD real-time evolution simulations on this model and analyze the photocurrent generation process numerically. 
By doing so, we demonstrate that the shift current arises from spinon excitations, the exotic excitations characteristic of the XXZ model.

The remainder of this paper is organized as follows. Section \ref{section2} details the model employed in our analysis, focusing on two key aspects: quantum spin excitations, whose crucial role in photocurrent generation will be confirmed later, and the phenomenologically introduced magnetoelectric coupling term required to model optical responses. Section \ref{section3} describes the real-time evolution calculations performed using the iTEBD method. The primary results of this work are presented in Sec.~\ref{section4}, which showcases the linear and second-order nonlinear optical conductivity spectra and demonstrates how these spectra reflect the excitation structure of the quantum spin system. 
In Sec.~\ref{section5}, we present a brief summary and discussions.

\section{Model} \label{section2}
To study the light-matter coupling in a spin system, we consider a 1D quantum spin chain model that incorporates a coupling with the external electric field through the exchange striction mechanism \cite{Fiebig_2005, Mostovoy}. The exchange striction arises from the electric polarization that depends on the spin configuration, and appears widely in multiferroic materials.

As a quantum spin chain, we consider the $S=1/2$ XXZ model,
\begin{equation}    \label{hamiltonian}
    \hat{H}_0=J\sum_j\left(\hat{S}_j^x\hat{S}_{j+1}^x+\hat{S}^y_j\hat{S}^y_{j+1}+\Delta\hat{S}^z_j\hat{S}^z_{j+1}\right).
\end{equation}
with antiferromagnetic interactions ($J>0$). In 1D systems, inherently strong quantum fluctuations tend to disturb the long-range order. 
It is well known that the XXZ model Eq.~\eqref{hamiltonian} is gapless for $|\Delta|<1$ \cite{Takahashi_1999,LMK}, 
while it is gapped for $\Delta>1$. 
In this paper, we focus on the gapped regime $\Delta>1$ that exhibits spontaneous symmetry breaking due to the presence of N\'eel order, making the system well suited for investigating nonreciprocal phenomena. We employ the numerical method utilizing matrix product states (MPSs), which demonstrates high accuracy for gapped systems. 

\begin{figure}
    \centering
    \includegraphics[width=0.6\linewidth]{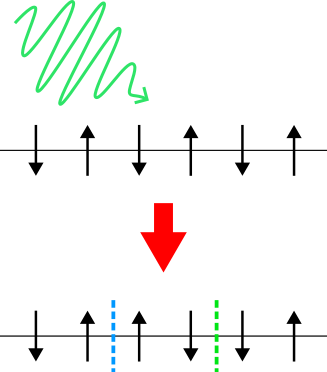}
    \caption{A schematic picture of spinon excitations in a spin chain. Spinons represent domain wall excitations in the Neel order.}
    \label{spinon}
\end{figure}

In the 1D XXZ model with $\Delta\gg 1$, the low-energy spin excitations in the N\'eel ordered phase are given by spinons \cite{FADDEEV1981375, LMK}, not by conventional magnons that correspond to spin waves \cite{Bloch1930,PhysRev.86.694,PhysRev.87.568}. Spinons are domain-wall excitations of the N\'eel-order and constitute an elementary excitation (Fig.~\ref{spinon}). In the N\'eel ordered phase, spin excitations manifest themselves as spin flips, creating pairs of domain walls within the N\'eel order. In 1D quantum spin chains, excitations of long-range domain-wall pairs become permissible, where the N\'eel order between the walls is inverted relative to that outside. The XY term $\hat{S}_j^x\hat{S}_{j+1}^x+\hat{S}^y_j\hat{S}^y_{j+1}=\frac{1}{2}(\hat{S}_j^+\hat{S}_{j+1}^-+\hat{S}^-_j\hat{S}^+_{j+1}$)  in the Hamiltonian Eq.~\eqref{hamiltonian} provokes the spin-pair flipping and works as a kinetic term of spinons. In the magnon picture, such a state with domain walls requires many magnons proportional to the distance between the walls, indicating that such elementary excitations are not captured well with the magnon-based description. 
These domain-wall excitations, spinons, are often interpreted as fractionalized (in this case, a magnon is fractionalized into two spinons) excitations resulting from the deconfinement of magnons and frequently exhibit fermionic behavior \cite{FADDEEV1981375,LMK}.

\begin{figure}
\centering
\includegraphics[width=0.7\linewidth]{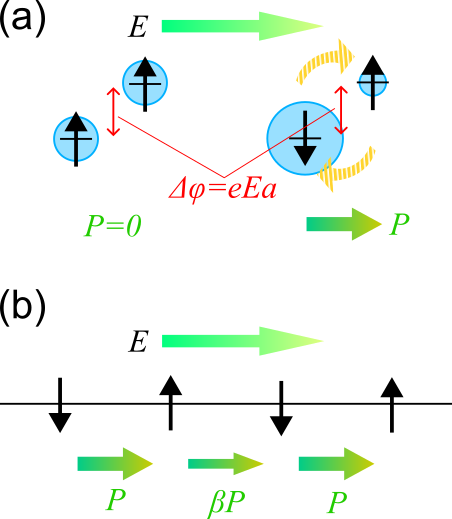}
\caption{(a) Schematics of the exchange striction mechanism for electric polarization in a spin system. The coupling between the electric field $E$ and the spin system arises from the modulation of the exchange interaction, induced by a potential gradient in the virtual hopping process. 
(b) A 1D multiferroic quantum spin model [Eq.~\eqref{polarization}], where spin-configuration-dependent electric polarization is generated on each bond due to the exchange striction mechanism and the broken inversion symmetry of the underlying electronic system.}
\label{Schematics}
\end{figure}

For a coupling between spin degrees of freedom and the electric field, we focus on the electric-field-induced modulation of the exchange interaction, as illustrated in Fig.~\ref{Schematics} \cite{Tokura_2014,PhysRevB.104.075139}. 
To see this, let us consider a two-electron system under strong on-site repulsive interactions, where electrons are localized at each site and an antiferromagnetic spin correlation arises due to an exchange process. 
Applying a potential difference between the two sites induces an asymmetry in electron density through the exchange process, i.e, electric polarization.
Since a virtual electron hopping between two sites is permitted only for antiferromagnetic spin configurations, 
the induced electric polarization through this process is proportional to the symmetric spin product $\boldsymbol{S}_i\cdot\boldsymbol{S}_j$. Generally, the mechanism by which electric polarization arises from the spin configuration via this inner-product form, thereby coupling the spin system to the electric field, is known as exchange striction. In Appendix \ref{append-1}, we derive this form of the coupling term starting from a concrete electronic model. Also, it should be noted that there is another microscopic mechanism of magnetoelectric coupling, the so-called Katsura-Nagaosa-Balatsky (KNB) mechanism \cite{PhysRevLett.95.057205,PhysRevLett.98.027203}. While the KNB mechanism is also a plausible candidate for inducing electric polarization, it inherently relies on spin-orbit coupling. Assuming that spin-orbit coupling gives a minor contribution, having light-element systems in mind, we perform calculations focusing on the exchange-strictionlike coupling, which is expected to provide the dominant contribution to the BPVE.  A similar BPVE result is also expected with the KNB mechanism.
In addition to these mechanisms, the magnetic field component of the light also induces spin dynamics through the Zeeman coupling. It was previously shown in Ref.~\cite{Ogino2024} that the magnetic component's contribution is smaller than that of the electric field component and that these two contributions exhibit distinct polarization dependencies, allowing them to be distinguished experimentally. Therefore, in the present analysis, we focused on the spin dynamics induced by the electric field component of light.

To the above XXZ model $\hat{H}_0$ [see Eq.~\eqref{hamiltonian}], we introduce a coupling between an electric field $E(t)$ and the polarization operator $\hat{P}$ in the exchange striction mechanism as,
\begin{gather}
    \label{hamiltonian_ex}  \hat{H}=\hat{H}_0-E(t)\hat{P} \\
    \label{polarization}    \hat{P}=\Pi\sum_j\beta_j\!\left(\hat{S}_j^x\hat{S}_{j+1}^x+\hat{S}^y_j\hat{S}^y_{j+1}+\Delta'\hat{S}^z_j\hat{S}^z_{j+1}\right) \\
	\left(\beta_{2j}=1,\,\beta_{2j+1}=\beta\right)
\end{gather}
Assuming that the exchange interaction possesses Ising-type uniaxial anisotropy prior to the application of an electric field, the interaction modulated via the exchange-striction mechanism is expected to also retain the same type of anisotropy. Therefore, throughout this paper, we set $\Delta'=\Delta$. The parameter $\beta$ controls the breaking of the site-center inversion symmetry. We remark that the bond-center inversion symmetry is spontaneously broken due to the N\'eel order. We note that $\Delta$ ($>1$) serves as a tuning parameter away from the quantum criticality at $\Delta=1$. Also, we set $\Pi/J=1$, for simplicity. 

To study the charge response of the quantum spin chain, one can focus on the displacement current, that is, the time derivative of polarization density. 
Using the Heisenberg equation of motion, the displacement current operator is defined as
$\hat{j}=\frac{1}{i\hbar N}\!\left[\hat{P},\hat{H}\right].$ 
By explicitly evaluating this operator for our model Hamiltonian and the polarization operator in Eq.~\eqref{polarization}, we obtain $\hat{j}=\hat{j}_1+\hat{j}_2$, where $\hat{j}_1$ and $\hat{j}_2$ have different dependences on $\Delta$ as
\begin{gather}
	\label{current_1}  \hat{j}_1=\frac{(1-\beta)\Pi J}{N}\sum_j(-1)^j\left(\hat{\boldsymbol{S}}_{j-1}\times\hat{\boldsymbol{S}}_{j+1}\right)^z\hat{S}^z_{j}, \\
	\label{current_2}  \hat{j}_2=\frac{\Delta(1-\beta)\Pi J}{N}\sum_j(-1)^j\!\sum_{\alpha=x,y}\!\left(\hat{\boldsymbol{S}}_{j-1}\times\hat{\boldsymbol{S}}_{j+1}\right)^\alpha\hat{S}^\alpha_{j}.
\end{gather}
This expression indicates that the current operator $\hat{j}$ looks similar to the scalar spin chirality   defined as a triple product of spins.
Here $j_2$ represents the in-plane components of $\hat{\boldsymbol{S}}_j$ and $\hat{\boldsymbol{S}}_{j-1}\times\hat{\boldsymbol{S}}_{j+1}$ and is enhanced by a factor of $\Delta$ compared to the $z$ component in $j_1$, which originates from the Ising anisotropy of the XXZ model [Eq.~\eqref{hamiltonian}] and the polarization operator [Eq.~\eqref{polarization}].
We note that a similar relationship between the scalar spin chirality and the current is known for electronic systems in the Mott phase \cite{PhysRevB.78.024402}.

\section{iTEBD simulation of optical processes}   \label{section3}
In this section, we study the optical responses of a quantum spin chain in terms of Eq.~\eqref{hamiltonian_ex} using iTEBD, which is an efficient algorithm to obtain the ground state and time evolution of 1D quantum many-body systems by representing a state as a tensor network~\cite{PhysRevLett.98.070201}. Numerical calculations are equipped by utilizing the ITensor library~\cite{10.21468/SciPostPhysCodeb.4}.

\begin{figure*}
    \centering
    \includegraphics[keepaspectratio,width=\linewidth]{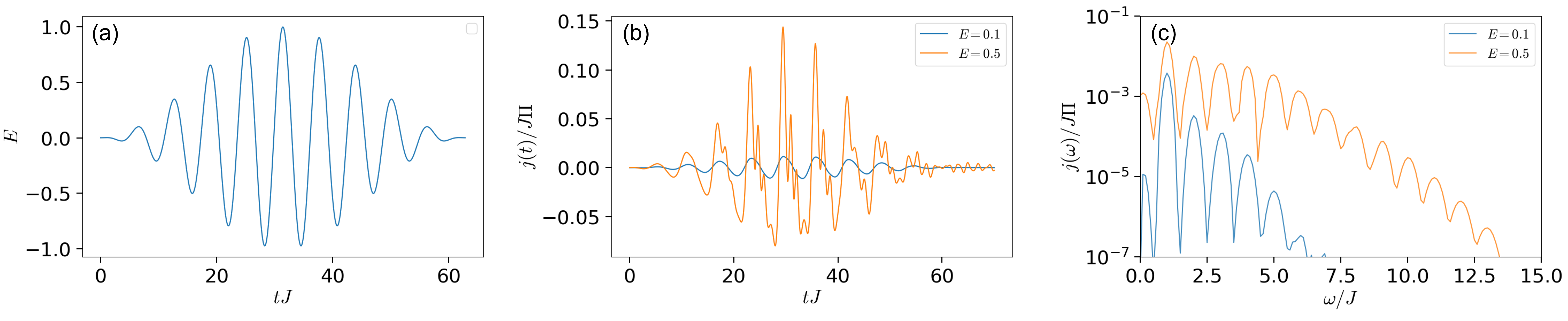}
    \caption{An example of the time-evolution simulations performed in this study. (a) Waveform of the laser electric field $E(t)$ ($E=1.0$, $\Omega/J=1.0$, $N_\mathrm{cyc}=10$). (b) Time evolution of the current $\langle j(t)\rangle$ ($\Delta=5$, $\beta=0.0$, $\Omega/J=1.0$, $N_\mathrm{cyc}=10$). (c) Frequency spectrum of $\langle j(t)\rangle$ ($\Delta=5$, $\beta=0.0$, $\Omega/J=1.0$, $N_\mathrm{cyc}=10$). The capture of current dynamics, including DC and HHG signatures as shown in these figures, indicates the capability of the iTEBD to handle non-perturbative regimes.}
    \label{profile_plot}
\end{figure*}

First, the ground state $\left|\Psi_\mathrm{GS}\right\rangle$ of $\hat{H}_0$ was obtained through the imaginary-time evolution method within the iTEBD framework and prepared as the initial state for the time evolution under the external electric field to study the optical responses. 
The real-time evolution was computed as
\begin{gather}
    \left|\Psi(t)\right\rangle=\hat{U}^\dagger(t)\!\left|\Psi_\mathrm{GS}\right\rangle \\
    \hat{U}^\dagger(t)=T\mathrm{exp}\!\left[-i\int_0^t\mathrm{d}t'\hat{H}(t')\right]
\end{gather}
for the time-dependent Hamiltonian $\hat{H}(t)$, where $T$ represents time-ordering. The electric field $E(t)$ is given by
\begin{align}    \label{laser-pulse}
E(t)&=E\sin^2\!\left(\frac{\Omega t}{2N_\mathrm{cyc}}\right)\cos{\Omega t},
\end{align}
as shown in in Fig.~\ref{profile_plot}(a).
Through this procedure, the current response is obtained by
\begin{equation}
    \left\langle j(t)\right\rangle=\left\langle\Psi(t)\right|\!\hat{j}\!\left|\Psi(t)\right\rangle.
\end{equation}
Representative profiles of these quantities are presented in Fig.~\ref{profile_plot}(b).

We perform a discrete Fourier transform for $\left\langle j(t)\right\rangle$ to obtain the current response in the frequency domain as shown in Fig.~\ref{profile_plot}(c).
The appearance of high harmonics  $\tilde{j}(\omega=n\Omega)$ in Fig.~\ref{profile_plot}(c) clearly shows that our iTEBD calculations well capture the nonlinear dynamics of the current response. When the external field amplitude is small ($E=0.1$ in Fig.~\ref{profile_plot}), the response intensity of high harmonics at $\omega=n\Omega$ is almost proportional to $E^n$, indicating that the $n$th harmonic $\tilde{j}(n\Omega)$ is generated by $n$ photon absorption process within the perturbative regime. On the other hand, for a large external field amplitude ($E=0.5$ in Fig.~\ref{profile_plot}), nonperturbative effects set in, as characterized by a plateau structure in the range of approximately $\omega/J\in[3,7]$, followed by a cutoff structure where the intensity rapidly decays at higher frequencies. The width of this plateau corresponds to the energy band of spinon excitations, which are the exact elementary excitations in the 1D XXZ model (see Sec.~\ref{section4} for details). These observations are consistent with the behavior of HHG in quantum spin systems found in a previous study on HHG in a quantum spin chain \cite{PhysRevB.99.184303,PhysRevB.100.214424}.

\section{Linear and Nonlinear Photoconductivity}    \label{section4}
In this section, we study the linear and second-order current responses of the quantum spin chain.

    \begin{figure*}
        \centering
        \includegraphics[keepaspectratio,width=0.9\linewidth]{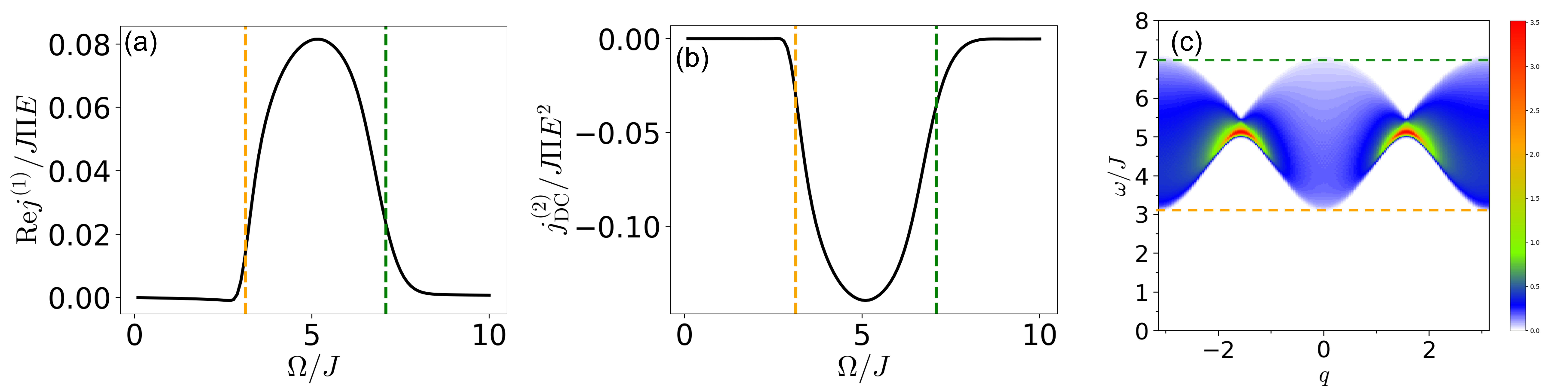}
        \caption{Numerical results of (a)(b)linear and nonlinear DC photoconductivity spectra ($\beta=0.0,\,E=10^{-2}$) and (c)the imaginary part of the spin susceptibility $\chi_{xx}(k,\omega)$ of 1D XXZ model [Eq.~\eqref{hamiltonian}]($\Delta=5$)。}
        \label{current_spectrum}
    \end{figure*}

For sufficiently weak laser intensity, the first-order current response $j^{(1)}$ is given by $\langle\tilde{j}(\Omega)\rangle$ as 
\begin{equation}
    j^{(1)}(\Omega)=\langle\tilde{j}(\Omega)\rangle
    =\sigma^{(1)}(\Omega)E(\Omega).
\end{equation}
Similarly, the second-order DC current response $j_\mathrm{DC}^{(2)}$ is given by the DC component $\langle\tilde{j}(\omega=0)\rangle$ as
\begin{equation}
    j^{(2)}_\mathrm{DC}=\langle\tilde{j}(\omega=0)\rangle
    =\sigma^{(2)}_\mathrm{DC}(\Omega)\left|E(\Omega)\right|^2.
\end{equation}
Therefore, we performed iTEBD simulations with weak electric fields, while keeping the field amplitude fixed at $E=10^{-2}\ll 1$ and varying the laser frequency $\Omega$ to obtain $\sigma^{(1)}(\Omega)$ and $\sigma^{(2)}_\mathrm{DC}(\Omega)$. 
Figure~\ref{current_spectrum}(a) and \ref{current_spectrum}(b) show the spectra of $\sigma^{(1)}(\Omega)$ and $\sigma^{(2)}_\mathrm{DC}(\Omega)$ obtained from measuring $\langle\tilde{j}(\Omega)\rangle$ and $\langle\tilde{j}(\omega=0)\rangle$, respectively.

The real part of the linear optical conductivity $\sigma^{(1)}(\omega)$ corresponds to the absorption coefficient. The result for $\mathrm{Re}\tilde{J}(\omega)\simeq\mathrm{Re}\sigma^{(1)}(\omega)E(\omega)$ shown in Fig.~\ref{current_spectrum}(a) exhibits nonzero optical absorption in the range $\omega/J\in[3,7]$. 
The origin of this optical absorption is attributed to spinon-pair excitations.
To verify this, we consider its correspondence with the dynamical spin susceptibility. It is given as the Fourier transform of the retarded correlation function
\begin{equation}
    \chi_{xx}(r,t)\equiv-i\left\langle\left[\hat{S}^x_r(t),\hat{S}^x_0\right]\right\rangle\Theta(t),
\end{equation}
and contains information on the density of states of the spinon-pair excitations, where $\Theta(t)$ represents the step function. 
The nonzero $\mathrm{Im}\chi_{xx}(k,\omega)$ in Fig.~\ref{current_spectrum}(c) corresponds to the region where spinon-pair excitations exist.
We can compute $\chi_{xx}(k,\omega)$ using TEBD efficiently and show the result for $\mathrm{Im}[\chi_{xx}(k,\omega)]$ in the color plot [Fig.~\ref{current_spectrum}(c)].
The vertical dotted lines in Fig.~\ref{current_spectrum} indicate the energy range of spinon excitations for the 1D XXZ model Eq.~\eqref{hamiltonian}, which coincides with the analytic result for the spinon-pair continuum $\omega/J \in [3.12, 7.08]$ obtained from the Bethe ansatz~\cite{PhysRevB.57.11429}.
It is clearly seen that the linear conductivity exhibits peaks within this spinon-pair continuum [Fig.~\ref{current_spectrum}(a)].

Next, let us look at the results for the nonlinear optical conductivity $\sigma_\mathrm{DC}^{(2)}(\omega)$ shown in Fig.~\ref{current_spectrum}(b). We find that a non-zero photocurrent is also generated in the energy region where optical absorption occurs. 
Since this absorption range corresponds to the excitation continuum of spinon pairs, this result indicates that the optical excitation of spinon pairs also induce photocurrent generation. 

Generally, the second-order DC current consists of several microscopic mechanisms including the shift current, the injection current and the ballistic current~\cite{Baltz,Sipe,Young-Rappe,doi:10.1126/sciadv.1501524,Resta24}. In particular, the latter two require the generation of free charge carriers for sustaining photocurrent. In our current setup, we consider optical excitations with photon energies below the electronic band gap, where charge carriers cannot be generated and no photocurrent arises from the injection or ballistic current mechanism. 
Instead, the optical excitation of spinon pairs generates a photocurrent via the shift current mechanism here. 
This can be interpreted as follows: (i) the spinon pairs possess electric polarization due to the broken inversion symmetry and the exchange striction; (ii) the change in the system's electric polarization upon the optical excitation of these pairs leads to DC generation $j=dp/dt$ through the shift current mechanism. 
In this sense, the second order DC current response $\sigma_\mathrm{DC}^{(2)}(\omega)$ in the present model constitutes the ``spinon shift current'' which is an extension of the shift current to emergent spinon excitations in a multiferroic quantum spin chain.

   \begin{figure*}
        \centering
        \includegraphics[keepaspectratio,width=\linewidth]{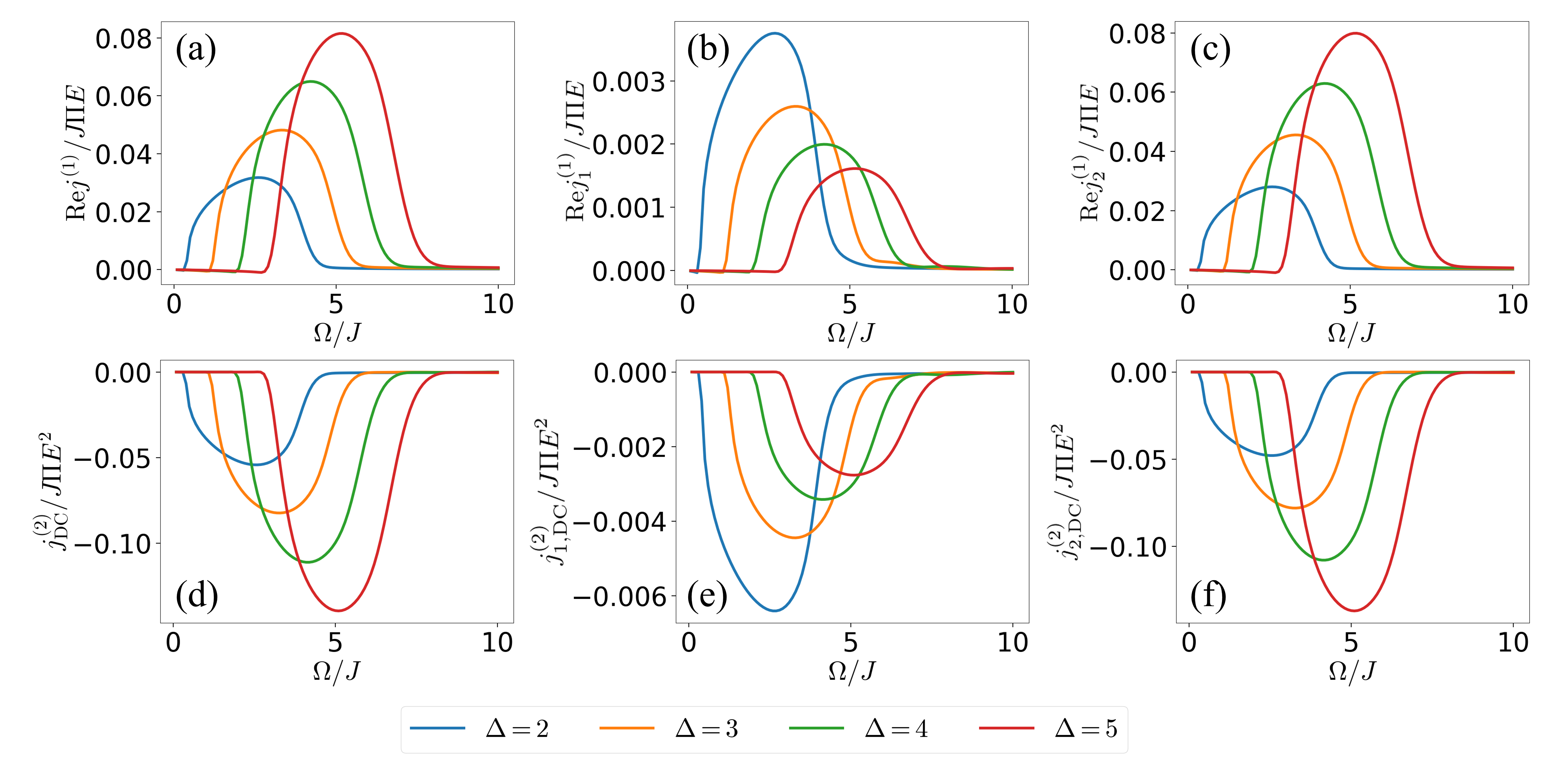}
        \caption{Laser frequency dependence of the current responses $\mathrm{Re}\langle\tilde{j}(\Omega)\rangle$ and $\langle\tilde{j}(\omega=0)\rangle$ for various values of a criticality parameter $\Delta$. While (a) and (d) are based on the dynamics of the expectation value of total current $\hat{j}$, (b) and (e) are those measured solely for $\hat{j}_1$ [Eq.~\eqref{current_1}], and (c) and (f) for $\hat{j}_2$ [Eq.~\eqref{current_2}].}
        \label{current_spectrum_comparison}
    \end{figure*}

We also performed simulations by varying the criticality parameter $\Delta$. 
Figures~\ref{current_spectrum_comparison}(a) and \ref{current_spectrum_comparison}(d) show the results of linear current response $\langle\tilde{j}(\Omega)\rangle$ and second-order DC current response $\langle\tilde{j}(\omega=0)\rangle$, respectively, for various values of $\Delta =2, 3, 4, \text{and}\, 5$.
We find that the peak structure is shifted with changing $\Delta$, which reflects the fact that the two-spinon excitation takes place around $\Omega \simeq \Delta$.
While Figs.~\ref{current_spectrum_comparison}(a) and \ref{current_spectrum_comparison}(d) are obtained from the expectation value of the total current operator $\hat{j}$ defined as the sum of $\hat{j}_1$ [Eq.~\eqref{current_1}] and $\hat{j}_2$ [Eq.~\eqref{current_2}], the remaining plots in Fig.~\ref{current_spectrum_comparison} are from only the component of $\hat{j}_1$ [(b) and (e)] and $\hat{j}_2$ [(c) and (f)]. One can see that the $\hat{j}_1$ response increases as $\Delta$ is reduced from 5 to 2, i.e., the system approaches the quantum critical point $\Delta=1$ [Figs.~\ref{current_spectrum_comparison}(b) and \ref{current_spectrum_comparison}(e)].
This behavior contrasts with $\hat{j}_2$, which is almost proportional to $\Delta$ [Figs.~\ref{current_spectrum_comparison}(c) and \ref{current_spectrum_comparison}(f)]. 
This trend is attributed to the fact that the scalar spin chirality
$
    (\hat{\boldsymbol{S}}_{j-1}\times\hat{\boldsymbol{S}}_{j+1})\cdot\hat{\boldsymbol{S}}_{j},
$
driving the current response, becomes more isotropic as the system moves from the strong $S_z$-anisotropy regime ($\Delta=5$) toward the weak regime ($\Delta=2$). This is consistent with the physical picture that the antiferromagnetic order becomes more isotropic as the system approaches the Heisenberg point $\Delta=1$.
Moreover, in the plots for the $\hat{j}_1$ component [Figs.~\ref{current_spectrum_comparison}(b) and \ref{current_spectrum_comparison}(e)], we can observe that those current responses exhibit an overall exponential increase, rather than a linear one, as $\Delta$ approaches the quantum critical point (QCP) at $\Delta=1$.  Since the in-plane spin susceptibility undergoes a critical divergence at the QCP,  the quantum criticality similarly leads to a critical enhancement in both the linear and the nonlinear current responses. In particular, it is interesting that the shift current can also exhibit critical divergence through the mediation of spinons. Although the decomposition of the current operator, motivated by its relation to the scalar spin chirality, may be difficult to isolate individually in experiments, it indicates that the total optical conductivity shows a critical enhancement driven by the in-plane component.

Finally, the ratio between $\sigma^{(1)}$ and $\sigma_\mathrm{DC}^{(2)}$ allows us to estimate the electric dipole moment carried by spinon pairs during the photovoltaic process. The electric dipole moment per spinon excitation, $p(\omega)$, serves as a physical quantity corresponding to the shift vector, a geometric quantity in the standard electron-hole shift current picture. However, this $p(\omega)$ cannot be directly expressed in terms of quantum geometric quantities of the spin system, such as the spin Berry connection. As analyzed in Appendix \ref{append-2}, the magnitude of the dipole moment remains approximately constant regardless of $\Delta$ within the range of $\Delta\in[2,5]$. 
While no significant change in the dipole moment is observed at least within this parameter range, there is a possibility of distinct behavior emerging as the system approaches the quantum critical point at $\Delta=1$, although this is numerically hard to access.

\section{Summary  and Discussion}    \label{section5}
We have investigated the role of spin excitations in the charge current responses in a multiferroic quantum spin chain via the iTEBD method. 
We found that the spinon excitations in a 1D quantum spin chain generate a photovoltaic effect, which can be interpreted as a spinon shift current. 
Namely, the BPVE occurs in the absence of free charge carriers and is driven by the electric polarization of the spinons, extending the conventional shift current mechanism for interband electronic excitations. 
Furthermore, a spinon can be viewed as an excitation resulting from the fractionalization of a magnon excitation in the spin wave theory; in this sense, the spinon shift current represents a photocurrent carried by fractionalized excitations in quantum spin systems. 
Since various fractionalized excitations are known to emerge in quantum spin systems, particularly in frustrated systems \cite{Savary_2017}, the possible shift current generation triggered by those exotic excitations is an interesting open question.

Previously, the shift current response induced by the magnon excitations was studied under the linear magnon approximation  \cite{PhysRevB.104.075139}.
Application of a similar analysis to our model [Eqs.~\eqref{hamiltonian}, \eqref{hamiltonian_ex} and \eqref{polarization}] within spin wave theory leads to the second-order DC response, where the two-magnon excitations contribute to an excitation peak located at $\Omega/J\sim 2\Delta$. 
Meanwhile, we have shown that the second-order DC conductivity exhibits an excitation peak at $\Omega/J\sim \Delta$ (see Fig.~\ref{current_spectrum_comparison}) from the iTEBD simulation. 
As detailed in Appendix \ref{append-3}, this discrepancy arises from the fact that the magnon expansion no longer accurately reflects the underlying excitation structure in low-dimensional quantum spin systems;
instead, their fractionalized counterparts, the spinons, dominate excitation processes such as responses to external fields. 
Our iTEBD simulation clearly demonstrates that such exotic excitations in low-dimensional quantum spins can indeed drive photocurrent generation in strongly correlated electron systems.

Note that we focus on the gapped region ($\Delta > 1$) in our iTEBD analysis, because the MPS representation is not very efficient in the gapless regime and the iTEBD simulation becomes inaccurate.

Finally, we estimate the orders of magnitude of the linear and nonlinear optical conductivity due to the spinon excitations. 
In the following, we assume the lattice constant $a\sim 3\,\text{\AA}$, the exchange energy $J\sim 10\,\mathrm{meV}$ and the on-site Coulomb energy $U\sim 5\,\mathrm{eV}$ \cite{10.1098/rspa.1963.0204, Phillips_2012}. 
The characteristic laser frequency is in the terahertz range as
$\Omega/2\pi\sim J/h\sim 3\times 10^{12}\,\mathrm{s}^{-1}=3\,\mathrm{THz}$. 

Using the relationship $\Pi E/J=ea E/U$ from the microscopic picture of electronic systems (see Appendix \ref{append-1}), the parameter $E=10^{-2}$ employed in our numerics assuming $\Pi/J=1$ corresponds to an electric field of $E=10^{-2}\times(U/ea)\sim 2\,\mathrm{MV}/\mathrm{cm}$, which is experimentally accessible for terahertz pulses \cite{adom.201900681}. 
In addition, $\Pi=ea(J/U)$ serves as a measure of the dipole strength in the spin system, which scales with the factor $J/U\sim 2\times 10^{-3}$ compared to that for electronic interband transitions.

The dimension of the current is recovered by the factor $J\Pi/\hbar=(ea/\hbar)(J^2/U)$. From the calculated results of the first-order current response (we choose the peak values of the $\Delta=2$ case), the order of the optical conductivity is estimated as in 
\begin{align*}
    \mathrm{Re}\,\sigma^{(1)}
    &\simeq 3\times 10^{-2}\times\frac{1}{a}\frac{e^2}{\hbar}\!\left(\frac{J}{U}\right)^2 
    \simeq 2\times10^{-3}\,\mathrm{A/V\cdot m}
\end{align*}
Note that we have normalized the current by the cross-sectional area of the chain, $a^2$, assuming that the current flows in a stack of one-dimensional chains. 
Similarly, the order of the second-order DC optical conductivity, derived from the calculated second-order DC response, is given by
\begin{align*}
    \sigma^{(2)}_\mathrm{DC} 
    &\simeq 5\times 10^{-2}\times\frac{e^3}{\hbar}\frac{J^2}{U^3} 
    \simeq 2\times 10^{-1}\,\mathrm{pA/V^2}
\end{align*}

In the experimental study of the photovoltaic effect mediated by spin excitations via terahertz light \cite{Ogino2024}, the photovoltaic response is evaluated using the Glass coefficient $G$ defined by $J_\mathrm{DC}=Gi_\mathrm{abs}w$, where $J_\mathrm{DC}$ is the second-order DC current, $i_\mathrm{abs}$ is the energy density absorbed from the pump light, and $w$ is the irradiated area of the sample. 
Since the absorbed energy is proportional to the real part of the linear conductivity $\mathrm{Re}\sigma^{(1)}$ and the square of the electric field amplitude, and the DC current is proportional to $\sigma^{(2)}_\mathrm{DC}$ and the square of the field amplitude, $G$ is given by the ratio of the two conductivities as $G=\sigma^{(2)}_\mathrm{DC}/\mathrm{Re}\sigma^{(1)}$ (The area of the irradiated surface cancels out with that of the photocurrent-carrying surface. Therefore, this equality requires the isotropy of the sample size). Consequently, the order of magnitude for $G$ derived from the numerical calculations in this paper is found to be $G\sim 1\times10^{-8}\,\mathrm{cm}/\mathrm{V}$. This value is just comparable to that of the magnon shift current in multiferroics $\mathrm{Eu}_{0.55}\mathrm{Y}_{0.45}\mathrm{MnO}_3$ reported in Ref.~\cite{Ogino2024}.

\begin{acknowledgements}
We thank Sota Kitamura and Youichi Yanase
for fruitful discussions.
This work was supported by JSPS KAKENHI, Grants No. 24K06891, 24H00191, 26K00662 (ST), 24H02231, 23K25816, 23K17665, 24K00568 (TM).
\end{acknowledgements}

\appendix

\section{Details of numerical analysis}
\begin{figure}
        \centering
        \includegraphics[keepaspectratio,width=0.8\linewidth]{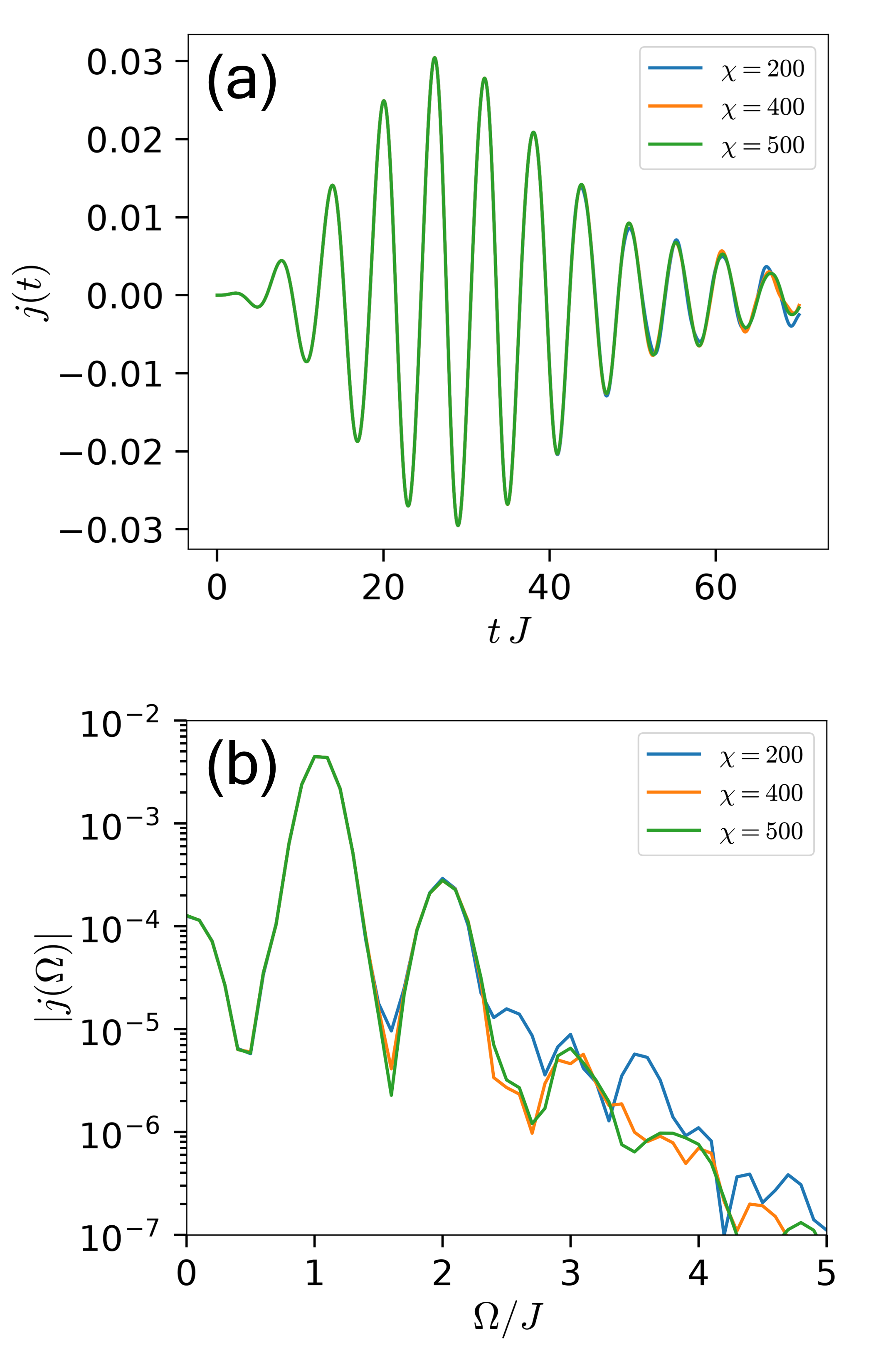}
        \caption{Numerical convergence with respect to the bond dimension $\chi$ in the iTEBD calculations. Each dynamics is calculated in the same way as Fig.\ref{profile_plot}(b) ($\Delta=2,\beta=0.0,E=0.1$). (a) Time evolution of the electric current $j(t)/J\Pi$. (b) HHG spectra $|j(\Omega)|/J\Pi$.}
        \label{bond-convergence}
    \end{figure}

To confirm the numerical validity and accuracy of our iTEBD calculations, we evaluated the convergence of the results with respect to the cutoff bond dimension $\chi$ of the MPS. Figure~\ref{bond-convergence} presents the calculations of the current dynamics, performed as described in Sec. \ref{section3} at $\Delta=2,\beta=0.0, E=0.1$ [see eqs. \eqref{hamiltonian}-\eqref{polarization} and \eqref{laser-pulse}], for $\chi = 200, 400,\text{and}\,500$. While minor quantitative differences in the high-frequency regime ($\Omega/J > 2$) are visible when comparing $\chi = 200$ with larger bond dimensions, the results for $\chi = 400$ and $\chi = 500$ show a good agreement across the frequency range shown. This saturation confirms that a bond dimension of $\chi = 400$, the value used in all the calculation in this study, is sufficient to obtain quantitatively converged time dynamics.

\section{Microscopic Derivation of Exchange Striction}  \label{append-1}
In this section, we demonstrate that the electric polarization operator in the form of $\boldsymbol{S}_i\cdot\boldsymbol{S}_j$ can be derived from an underlying electronic system. 

\begin{widetext}
Let us consider a two-site, two-electron system, described by the Hamiltonian,
\begin{equation}
    \hat{H}(\boldsymbol{E})=\sum_{\sigma=\uparrow,\downarrow}\!\left[\mu\hat{c}^\dagger_{\sigma\!A}\hat{c}_{\sigma\!A}-\Delta(\boldsymbol{E})\hat{c}^\dagger_{\sigma\!B}\hat{c}_{\sigma\!B}\right]
    +t\sum_{\sigma=\uparrow,\downarrow}\!\left(\hat{c}^\dagger_{\sigma\!B}\hat{c}_{\sigma\!A}+\mathrm{h.c.}\right)+U\sum_{j}\!\left(\hat{n}_{\uparrow A}\hat{n}_{\downarrow A}+\hat{n}_{\uparrow B}\hat{n}_{\downarrow B}\right).
\end{equation}
Here $t$ is the hopping integral, $U$ is the on-site Coulomb energy, $\mu$ is the on-site potential difference and $\Delta(\boldsymbol{E})=-e\boldsymbol{a}\cdot\boldsymbol{E}$ is a potential gradient induced  by an electric field $\bm{E}$.
In the strong-coupling limit $U\gg t$, the effective Hamiltonian of the system on the restricted state space, where  each site is occupied with a single electron, is given by
\begin{equation}
    \hat{H}_\mathrm{eff}(\boldsymbol{E})=2t^2\!\left(\frac{1}{U+\mu-\Delta(\boldsymbol{E})}+\frac{1}{U-\mu+\Delta(\boldsymbol{E})}\right)\!\hat{\boldsymbol{S}}_A\cdot\hat{\boldsymbol{S}}_B
\end{equation}
Assuming the electric field gradient $\Delta(\boldsymbol{E})\ll U,\mu$ so as to maintain the localized electron picture, the Hamiltonian is expanded in $\bm{E}$ as
\begin{equation}
    \hat{H}_\mathrm{eff}(\boldsymbol{E})=\hat{H}_\mathrm{eff}(\boldsymbol{E}=\boldsymbol{0})+2et^2\boldsymbol{a}\cdot\boldsymbol{E}\!\left[\frac{1}{(U-\mu)^2}-\frac{1}{(U+\mu)^2}\right]\!\hat{\boldsymbol{S}}_A\cdot\hat{\boldsymbol{S}}_B+\mathcal{O}(E^2)
\end{equation}
Since the electric polarization operator $\hat{\boldsymbol{P}}$ is generally given by $\hat{\boldsymbol{P}}=-\partial\hat{H}/\partial\boldsymbol{E}$, it takes the form,
\begin{equation}
    \hat{P}_\mathrm{eff}\simeq-\frac{\partial\hat{H}_\mathrm{eff}}{\partial\boldsymbol{E}}=-2et^2\boldsymbol{a}\!\left[\frac{1}{(U-\mu)^2}-\frac{1}{(U+\mu)^2}\right]\!\hat{\boldsymbol{S}}_A\cdot\hat{\boldsymbol{S}}_B+\mathcal{O}(E^2)
\end{equation}
\end{widetext}

The above polarization operator vanishes in the absence of the onsite potential difference $\mu$,
reflecting the fact that the underlying electronic system must break inversion symmetry for generating a nonzero electric polarization. 

\section{Dipole moment per a spinon pair} \label{append-2}
To investigate the relationship between the changes in spinon excitations and the DC response as $\Delta$ varies, we focus on the following quantity:
\begin{equation}
    p(\omega)\equiv\hbar\omega\frac{\sigma^{(2)}_\mathrm{DC}(\omega)}{\mathrm{Re}\sigma^{(1)}(\omega)},
\end{equation}
which describes an effective electric dipole moment carried by the spinon excitations.
Specifically, in the process of the second-order DC response, the number of elementary excitations $r$ generated per unit time is obtained by dividing the work done by the current per unit time, $\boldsymbol{J}\cdot\boldsymbol{E}=\mathrm{Re}\sigma^{(1)}E^2$ 
by the photon energy $\hbar\omega$. 
Consequently, the quantity $p(\omega)$, obtained by dividing the DC component $\sigma^{(2)}_\textrm{DC}E^2$  (representing the total excited dipole moments) by the excitation rate $r$, can be regarded as the dipole moment associated with a single elementary excitation. 

\begin{figure}
        \centering
        \includegraphics[keepaspectratio,width=0.8\linewidth]{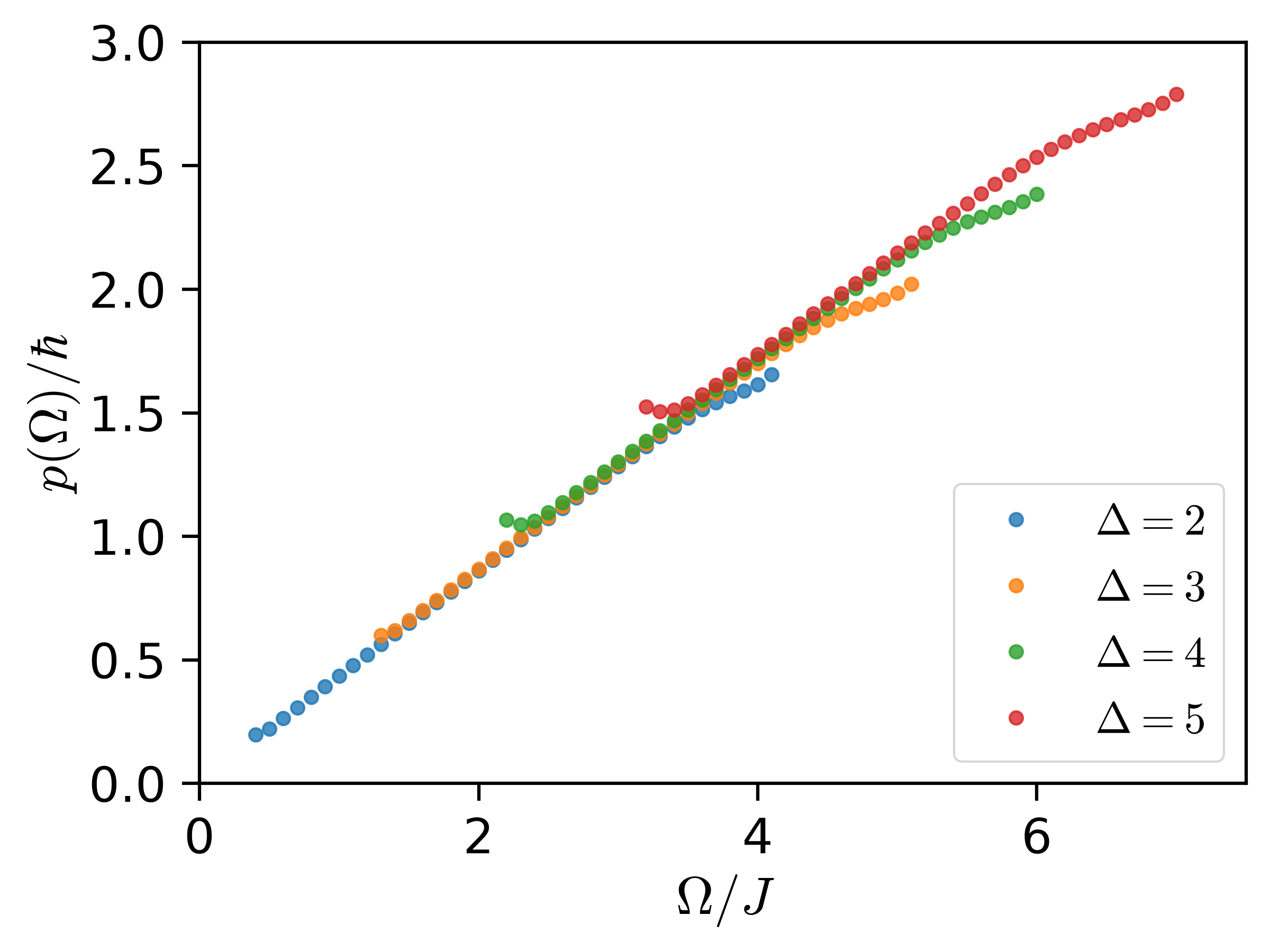}
        \caption{Dipole moment per a spinon pair $p(\Omega)\equiv \hbar\Omega\sigma^{(2)}_\mathrm{DC}(\Omega)/\sigma^{(1)}(\Omega)$ plotted as a function of $\Omega$. The data are shown for the energy window where the spinon excitation takes place, as shown in Fig.~\ref{current_spectrum_comparison}.}
        \label{spinon_pol}
    \end{figure}

Fig.~\ref{spinon_pol} shows the numerical result for $p(\omega)$ for each dataset shown in Fig.~\ref{current_spectrum_comparison}. 
We find that the behavior of $p(\Omega)$ remains nearly unchanged for different values of $\Delta$.
Whether this behavior persists as the system approaches the quantum critical point at $\Delta=1$ requires further investigation, which entails higher numerical cost, and remains an interesting future issue.

The Glass coefficient $G$ discussed in Sect.~\ref{section5} actually has a relationship with $p(\omega)$ as $p(\omega)=\hbar\omega G$. 
This allows for an estimation of the order of magnitude, yielding $p/e\sim 1\times 10^{-10}\,\mathrm{cm}\sim 10^{-2}\times a$ (Note that the peak frequency holds $\hbar\Omega\sim\Delta J$).

\section{Magnon and Jordan-Wigner fermion expansion}   \label{append-3}
In this section, we present theoretical analyses of the optical response in our model via two types of quasi-particle approximation methods: the magnon expansion and the Jordan-Wigner transformation.

A previous study \cite{PhysRevB.104.075139} calculated optical conductivity by mapping the spin system to a bosonic system via the Holstein-Primakoff transformation and approximating it as a free-particle system, truncating the Hamiltonian at the quadratic order of bosons (linear magnon expansion). However, this magnon expansion is generally not valid for systems with small spin $S$, low dimensionality, or antiferromagnetic interactions, owing to strong quantum fluctuations. To begin with, let us examine this limitation: Performing Holstein-Primakoff transformation (on the assumption of N\'eel-ordered ground state)
\begin{widetext}
            \begin{gather}
                \hat{S}_{2j}^+\equiv\hat{a}^\dagger_j\sqrt{2S-\hat{a}^\dagger_j\hat{a}_j},\quad
                \hat{S}_{2j}^-\equiv\sqrt{2S-\hat{a}^\dagger_j\hat{a}_j}\hat{a}_j,\quad
                \hat{S}_{2j}^z\equiv \hat{a}^\dagger_j\hat{a}_j-S \\
                \hat{S}_{2j+1}^+\equiv\sqrt{2S-\hat{b}^\dagger_j\hat{b}_j}\hat{b}_j,\quad
                \hat{S}_{2j+1}^-\equiv\hat{b}^\dagger_j\sqrt{2S-\hat{b}^\dagger_j\hat{b}_j},\quad
                \hat{S}_{2j+1}^z\equiv S-\hat{b}^\dagger_j\hat{b}_j
            \end{gather}
        to $\hat{H}_0, \hat{P}$ and $\hat{j}$ yields
            \begin{gather}
                \hat{H}_0=-NS^2\Delta+2S\Delta\sum_j\!\left(\hat{a}^\dagger_j\hat{a}_j+\hat{b}^\dagger_j\hat{b}_j\right)\!+S\sum_{j}\!\left(\hat{a}_j\hat{b}_j+\hat{a}^\dagger_j\hat{b}^\dagger_j+\hat{b}_j\hat{a}_{j+1}+\hat{b}^\dagger_j\hat{a}^\dagger_{j+1}\right) \\
                \hat{P}=\frac{(1+\beta)\Delta\Pi}{2}\!\left[-NS^2+2S\sum_j\!\left(\hat{a}^\dagger_j\hat{a}_j+\hat{b}^\dagger_j\hat{b}_j\right)\right]\!+\Pi S\sum_{j}\!\left[\hat{a}_j\hat{b}_j+\hat{a}^\dagger_j\hat{b}^\dagger_j+\beta\!\left(\hat{b}_j\hat{a}_{j+1}+\hat{b}^\dagger_j\hat{a}^\dagger_{j+1}\right)\right] \\
                \hat{j}=\frac{i(1-\beta)\Pi S^2}{2N}\sum_j\left(\hat{a}^\dagger_j\hat{b}^\dagger_{j}-\hat{a}_j\hat{b}_{j}+\hat{b}_j\hat{a}_{j+1}-\hat{b}^\dagger_j\hat{a}^\dagger_{j+1}\right)
            \end{gather}
        Here, $S\gg\langle\hat{a}^\dagger_j\hat{a}_j\rangle,\langle\hat{b}^\dagger_j\hat{b}_j\rangle$ and $J_{xy},J_z\gg h$ are additionally assumed. Due to the spatial periodicity of the system, $\hat{a}_j,\hat{a}_j^\dagger,\hat{b}_j,\hat{b}^\dagger_j$ can be applied Fourier transformation
            \begin{equation}
                \left\{
                    \begin{gathered}
                        \hat{a}_j=\sqrt{\frac{2}{N}}\sum_k\hat{a}_ke^{2ijk} \\
                        \hat{a}^\dagger_j=\sqrt{\frac{2}{N}}\sum_k\hat{a}^\dagger_ke^{-2ijk}
                    \end{gathered}
                \right. \qquad \left\{
                    \begin{gathered}
                        \hat{b}_j=\sqrt{\frac{2}{N}}\sum_k\hat{b}_ke^{i(2j+1)k} \\
                        \hat{b}^\dagger_j=\sqrt{\frac{2}{N}}\sum_k\hat{b}^\dagger_ke^{-i(2j+1)k}
                    \end{gathered}
                \right.
            \end{equation}
        Then, $\hat{H}_0, \hat{P}$ and $\hat{j}$ can be written as
            \begin{gather}
                \hat{H}_0=-NS^2\Delta-NS\Delta+2S\sum_{k}\hat{\Psi}^\dagger_k\!
                    \begin{pmatrix}
                        \Delta & \cos{k} \\
                        \cos{k} & \Delta
                    \end{pmatrix}
                \!\hat{\Psi}_k \\
                \hat{P}=-\frac{(1+\beta)N\Pi}{2}(S^2\Delta+S\Delta)+\Pi S\sum_k\hat{\Psi}^\dagger_k\!
                    \begin{pmatrix}
                        (1+\beta)\Delta & e^{-ik}+\beta e^{ik} \\
                        e^{ik}+\beta e^{-ik} & (1+\beta)\Delta
                    \end{pmatrix}
                \!\hat{\Psi}_k \\
                \hat{j}=\frac{(1-\beta)\Pi S^2}{N}\sum_k\hat{\Psi}^\dagger_k\!
                    \begin{pmatrix}
                        0 & \sin{k} \\
                        -\sin{k} & 0
                    \end{pmatrix}
                \!\hat{\Psi}_k
            \end{gather}
        where $\hat{\Psi}_k\equiv\left(\hat{a}_k,\,\hat{b}^\dagger_{-k}\right)$. Further, performing Bogoliubov transformation
            \begin{gather}
                \hat{\alpha}_k\equiv\hat{a}_k\cosh{\theta_k}-\hat{b}^\dagger_{-k}\sinh{\theta_k},\quad\hat{\beta}_{-k}=\hat{b}_{-k}\cosh{\theta_k}-\hat{a}^\dagger_k\sinh{\theta_k} \\
                \therefore\,
                    \begin{pmatrix}
                        \hat{\alpha}_k \\ \hat{\beta}^\dagger_{-k}
                    \end{pmatrix}
                =
                    \begin{pmatrix}
                        \cosh{\theta_k} & -\sinh{\theta_k} \\
                        -\sinh{\theta_k} & \cosh{\theta_k}
                    \end{pmatrix}
                \!\hat{\Psi}_k,\quad \hat{\Psi}_k=
                    \begin{pmatrix}
                        \cosh{\theta_k} & \sinh{\theta_k} \\
                        \sinh{\theta_k} & \cosh{\theta_k}
                    \end{pmatrix}
                \!
                    \begin{pmatrix}
                        \hat{\alpha}_k \\ \hat{\beta}^\dagger_{-k}
                    \end{pmatrix}
            \end{gather}
        such that $\hat{H}_0$ is diagonaized (i.e. $\tanh{2\theta_k}=-\cos{k}/\Delta$), 
            \begin{gather}
                \hat{H}_0=-N\Delta S(S-1)+2S\sum_k\hat{\Psi}^\dagger_kV^{-1}_k\!
                    \begin{pmatrix}
                        \displaystyle\gamma_k & 0 \\
                        0 & \displaystyle\gamma_k 
                    \end{pmatrix}
                \!V^{-1}_k\hat{\Psi}_k \\
                \hat{P}=-\frac{(1+\beta)N\Pi\Delta}{2}S(S-1)+\Pi S\sum_k\hat{\Psi}^\dagger_kV^{-1}_k\!
                    \begin{pmatrix}
                        (1+\beta)\gamma_k & -i(1-\beta)\sin{k} \\
                        i(1-\beta)\sin{k} & (1+\beta)\gamma_k
                    \end{pmatrix}
                \!V^{-1}_k\hat{\Psi}_k \\
                \hat{j}=\frac{(1-\beta)\Pi S^2}{N}\sum_k\hat{\Psi}^\dagger_kV^{-1}_k\!
                    \begin{pmatrix}
                        0 & \sin{k} \\
                        -\sin{k} & 0
                    \end{pmatrix}
                \!V^{-1}_k\hat{\Psi}_k
            \end{gather}
        where $\gamma_k\equiv\sqrt{\Delta^2-\cos^2k}$ and 
            \begin{equation}
                V_k\equiv
                    \begin{pmatrix}
                        \cosh{\theta_k} & \sinh{\theta_k} \\
                        \sinh{\theta_k} & \cosh{\theta_k}
                    \end{pmatrix}
                ,\quad V^{-1}_k\equiv
                    \begin{pmatrix}
                        \cosh{\theta_k} & -\sinh{\theta_k} \\
                        -\sinh{\theta_k} & \cosh{\theta_k}
                    \end{pmatrix}
                =
                    \begin{pmatrix}
                        1 & 0 \\
                        0 & -1
                    \end{pmatrix}
                \!V_k\!
                    \begin{pmatrix}
                        1 & 0 \\
                        0 & -1
                    \end{pmatrix}.
            \end{equation}

        From \cite{PhysRevB.104.075139}, the linear and DC conductivity are found to be obtained by
            \begin{gather}
                \label{linear-cond} \sigma^{(1)}(\omega)=-i\Omega\int_{-\pi}^{\pi}\frac{\mathrm{d}k}{2\pi}\sum_{ab}\widetilde{\Pi}_{ab}\widetilde{\Pi}_{ba}\frac{f_a-f_b}{\Omega-(\epsilon_{b}-\epsilon_{a})+i\delta} \\
                \label{2ndDC-cond}  \sigma^{(2)}_\mathrm{DC}(\Omega)=-i\int^{\pi}_{-\pi}\frac{\mathrm{d}k}{2\pi}\sum_{ab}\widetilde{\Pi}_{ab}\widetilde{\Pi}_{ba}(\widetilde{\Pi}_{aa}-\widetilde{\Pi}_{bb})\left(\frac{f_a-f_b}{\Omega-(\epsilon_a-\epsilon_b)+i\delta}+\frac{f_a-f_b}{-\Omega-(\epsilon_a-\epsilon_b)+i\delta}\right)
            \end{gather}
        where $\widetilde{\Pi}$ is $2\times 2$ matrix given by
            \begin{equation}
                \widetilde{\Pi}=\Pi S
                    \begin{pmatrix}
                        (1+\beta)\gamma_k & i(1-\beta)\sin{k} \\
                        i(1-\beta)\sin{k} & -(1+\beta)\gamma_k
                    \end{pmatrix}
            \end{equation}
        and $\epsilon_1=2S\gamma_k,\,\epsilon_2=-2S\gamma_k,\,f_a=\Theta(-\epsilon_a)$. Therefore, all the factors in the integrands are $k$-dependent. Calculating these formulae, we obtain
            \begin{equation}
                \sigma^{(1)}(\Omega)=i\Omega(1-\beta)^2\Pi^2S^2\int_{-\pi}^\pi\frac{\mathrm{d}k}{2\pi}\sin^2k\!\left(\frac{1}{\Omega-4S\gamma_k+i\delta}-\frac{1}{\Omega+4S\gamma_k+i\delta}\right)
            \end{equation}
            \begin{multline}
                \sigma_\mathrm{DC}^{(2)}(\Omega)=i(1-\beta)^2(1+\beta)\Pi^3S^3\int_{-\pi}^\pi\frac{\mathrm{d}k}{2\pi}\sin^2k\sqrt{\Delta^2-\cos^2{k}} \\
                \times\left(-\frac{1}{\Omega-4S\gamma_k+i\delta}+\frac{1}{\Omega+4S\gamma_k-i\delta}-\frac{1}{\Omega+4S\gamma_k+i\delta}+\frac{1}{\Omega-4S\gamma_k-i\delta}\right)
            \end{multline}
When $S=1/2$, $\mathrm{Im}\sigma^{(1)}$ and $\sigma^{(2)}_\mathrm{DC}$ consist of contributions from terms in the $k$-sum that satisfy
\begin{equation}
    \Omega=4S\gamma_k=2\sqrt{\Delta^2-\cos^2k}
\end{equation}
Thus, calculations based on the magnon expansion predict that the spectral peaks for both linear and nonlinear responses should be located around $\Omega/J\sim2\Delta$. This result, however, contradicts the findings obtained from our numerical simulations (see Fig.~\ref{current_spectrum}).

As previously discussed in Sec.~\ref{section2}, the magnon picture fails to capture the excitation structure of antiferromagnetic spin chains. The discrepancy between the analysis using the Holstein-Primakoff transformation and our numerical results stems solely from the inaccuracy of the former. On the other hand, it is  still possible to validate our numerical findings using the Green function formalism as demonstrated above, if one employs an alternative mapping to a fermion system via the Jordan-Wigner transformation
\begin{equation}
    \hat{S}_{j}^+\equiv\mathrm{exp}\!\left[i\pi\sum_{r=1}^{j-1}\hat{c}^\dagger_r\hat{c}_r\right]\hat{c}^\dagger_j,\quad
    \hat{S}_{j}^-\equiv\hat{c}_j\,\mathrm{exp}\!\left[-i\pi\sum_{r=1}^{j-1}\hat{c}^\dagger_r\hat{c}_r\right],\quad
    \hat{S}_{j}^z\equiv \hat{c}^\dagger_j\hat{c}_j-\frac{1}{2}
\end{equation}
This transformation does not need any underlying order, in contrast to the Holstein-Primakoff transformation, and this is why the Jordan-Wigner analysis is valid in 1D quantum spin chains. In particular, the exact solution of the 1D transverse-field Ising model can be derived from the fermion system obtained through this transformation \cite{LIEB1961407}. Furthermore, by approximating the resulting Hamiltonian of the fermion system in a quadratic form, one can perform calculations analogous to those in Ref.~\cite{PhysRevB.104.075139}. In the following, we discuss this approach using the Ising limit ($\Delta\to\infty$) as an illustrative example.

Reorienting the spin axis associated with Ising anisotropy to the $x$ direction and performing the Jordan-Wigner transformation, we obtain
\begin{gather}
    \frac{\hat{H}_0}{\Delta}\to J\sum_j\hat{S}^x_j\hat{S}^x_{j+1}=\frac{J}{4}\sum_j\left(\hat{c}^\dagger_j-\hat{c}_j\right)\!\left(\hat{c}^\dagger_{j+1}+\hat{c}_{j+1}\right),\\
    \frac{\hat{P}}{\Delta}\to\Pi\sum_j\beta_j\hat{S}^x_j\hat{S}^x_{j+1}=\frac{\Pi}{4}\sum_j\beta_j\!\left(\hat{c}^\dagger_j-\hat{c}_j\right)\!\left(\hat{c}^\dagger_{j+1}+\hat{c}_{j+1}\right)
\end{gather}
Furthermore, applying the Fourier transform
\begin{equation}
                \left\{
                    \begin{gathered}
                        \hat{c}_{2j+1}=\sqrt{\frac{2}{N}}\sum_k\hat{c}_{k,A}e^{2ijk} \\
                        \hat{c}^\dagger_{2j+1}=\sqrt{\frac{2}{N}}\sum_k\hat{c}^\dagger_{k,A}e^{-2ijk}
                    \end{gathered}
                \right. \qquad \left\{
                    \begin{gathered}
                        \hat{c}_{2j}=\sqrt{\frac{2}{N}}\sum_k\hat{b}_ke^{i(2j+1)k} \\
                        \hat{c}^\dagger_{2j}=\sqrt{\frac{2}{N}}\sum_k\hat{b}^\dagger_ke^{-i(2j+1)k}
                    \end{gathered}
                \right.
\end{equation}
one can get the following expressions (note the sublattice structure, which accounts for the N\'eel-ordered state):
\begin{gather}
    \hat{H}_0=\frac{J\Delta}{2}\sum_{k>0}\left.
                        \begin{pmatrix}
                            \hat{c}^\dagger_{k,A} \\
                            \hat{c}_{-k,A} \\
                            \hat{c}^\dagger_{k,B} \\
                            \hat{c}_{-k,B}
                        \end{pmatrix}
                    \!\!\right.^\mathrm{T}\!
                        \begin{pmatrix}
                            & & \cos{k} & i\sin{k} \\
                            & & -i\sin{k} & -\cos{k} \\
                            \cos{k} & i\sin{k} & & \\
                            -i\sin{k} & -\cos{k} & & 
                        \end{pmatrix}
                        \begin{pmatrix}
                            \hat{c}_{k,A} \\
                            \hat{c}^\dagger_{-k,A} \\
                            \hat{c}_{k,B} \\
                            \hat{c}^\dagger_{-k,B}
                        \end{pmatrix} \\
    \hat{\Pi}=\frac{\Pi\Delta}{4}\sum_{k>0}\left.
                        \begin{pmatrix}
                            \hat{c}^\dagger_{k,A} \\
                            \hat{c}_{-k,A} \\
                            \hat{c}^\dagger_{k,B} \\
                            \hat{c}_{-k,B}
                        \end{pmatrix}
                    \!\!\right.^\mathrm{T}\!
                        \begin{pmatrix}
                            & & e^{ik}+\beta e^{-ik} & e^{ik}-\beta e^{-ik} \\
                            & & -e^{ik}+\beta e^{-ik} & -e^{ik}-\beta e^{-ik} \\
                            \beta e^{ik}+e^{-ik} & \beta e^{ik}-e^{-ik} & & \\
                            -\beta e^{ik}+e^{-ik} & -\beta e^{ik}-e^{-ik} & &
                        \end{pmatrix}
                        \begin{pmatrix}
                            \hat{c}_{k,A} \\
                            \hat{c}^\dagger_{-k,A} \\
                            \hat{c}_{k,B} \\
                            \hat{c}^\dagger_{-k,B}
                        \end{pmatrix}
\end{gather}
Once the Bogoliubov–de Gennes-Bloch (BdG-Bloch) basis representations for these operators are established, the optical conductivities can be evaluated using a formalism nearly identical to the Green function approach employed in Ref.~\cite{PhysRevB.104.075139}. Specifically, the general expressions for $\sigma^{(1)}$ and $\sigma^{(2)}_\mathrm{DC}$ are given by Eqs.\eqref{linear-cond} and \eqref{2ndDC-cond}, respectively. The primary differences from Ref.~\cite{PhysRevB.104.075139} are that $\epsilon_a$ represents the eigenvalues of the Bloch Hamiltonian at wavevector $k$, and $\tilde{\Pi}$ is the representation of $\hat{\Pi}$ in the basis that diagonalizes the Bloch Hamiltonian of $\hat{H}_0$; these arise from the fact that a unitary transformation, rather than a Bogoliubov transformation, is sufficient to diagonalize the Bloch Hamiltonian in the fermion system. In the Ising limit, the eigenvalues of the Bloch Hamiltonian are $\epsilon_a=\pm J\Delta/2$, from which Eqs.~\eqref{linear-cond} and \eqref{2ndDC-cond} imply that the optical conductivity peaks appear at $\Omega/J=\Delta$. 

The above analytical results in the Ising limit actually imply that the quasiparticles of this system do not couple with the external electric field. 
Specifically, the matrix representation of $\hat{\Pi}$ in the BdG-Bloch basis (corresponding to $\tilde{\Pi}$ of the magnon case) commutes with the Hamiltonian as $[\hat\Pi, \hat H_0]=0$.
This forbids the spinon excitation by the external electric field, causing the optical conductivity calculated from the above formulas to vanish. 
Meanwhile, the XXZ model employed in our numerical calculations includes a weak in-plane interaction term $H_\mathrm{inplane}$ in addition to the Ising interaction. In the Jordan-Wigner fermion picture, the addition of $H_\mathrm{inplane}$ modifies the fermion eigenstates, allowing optical excitation of the spinons with $[\hat\Pi, \hat H_0]\neq 0$. Also, $H_\mathrm{inplane}$ introduces fermion interaction terms within a perturbative regime, leading to broadening of the spinon excitation spectrum. Thus, in the presence of the in-plane interaction $H_\mathrm{inplane}$, the optical conductivity is expected to show a peak structure around $\Omega=J\Delta$. 
This is consistent with the numerical results for conductivity shown in Fig.~\ref{current_spectrum_comparison}.
\end{widetext}

\bibliographystyle{apsrev4-1}
\bibliography{references}

\end{document}